\documentclass[aps,prb,twocolumn,superscriptaddress,amsmath,amssymb]{revtex4}

\usepackage{amssymb}
\usepackage{graphicx}
\usepackage{tabularx}
\usepackage{bm}
\usepackage{epsfig}
\usepackage[ansinew]{inputenc}
\usepackage{float}

\begin{document}

\title{The effect of structural and magnetic disorder on the 3$d$-5$d$ exchange interactions of La$_{2-x}$Ca$_{x}$CoIrO$_{6}$}

\author{L. Bufai\c{c}al}
\email{lbufaical@ufg.br}
\affiliation{Instituto de F\'{\i}sica, Universidade Federal de Goi\'{a}s, 74001-970, Goi\^{a}nia, GO, Brazil}

\author{E. Sadrollahi}
\affiliation{Institut f{\"u}r Physik der Kondensierten Materie, Technische Universit{\"a}t Braunschweig, 38106 Braunschweig, Germany}
\affiliation{Institut f{\"u}r Festk{\"o}rper- und Materialphysik, Technische Universit{\"a}t Dresden, 01069 Dresden, Germany}

\author{F. J. Litterst}
\email{j.litterst@tu-braunschweig.de}
\affiliation{Institut f{\"u}r Physik der Kondensierten Materie, Technische Universit{\"a}t Braunschweig, 38106 Braunschweig, Germany}
\affiliation{Centro Brasileiro de Pesquisas F\'{\i}sicas, 22290-180, Rio de Janeiro, RJ, Brazil}

\author{D. Rigitano}
\affiliation{Instituto de F\'{\i}sica ``Gleb Wataghin", UNICAMP, 13083-859, Campinas, SP, Brazil}

\author{E. Granado}
\affiliation{Instituto de F\'{\i}sica ``Gleb Wataghin", UNICAMP, 13083-859, Campinas, SP, Brazil}

\author{L. T. Coutrim}
\affiliation{Instituto de F\'{\i}sica, Universidade Federal de Goi\'{a}s, 74001-970, Goi\^{a}nia, GO, Brazil}
\affiliation{Laborat{\'o}rio Nacional de Luz S\'{\i}ncrotron, Centro Nacional de Pesquisa em Energia e Materiais, 13083-970, Campinas, SP, Brazil}

\author{E. B. Ara\'{u}jo}
\affiliation{Department of Physics and Chemistry, S\~{a}o Paulo State University, 15385-000 Ilha Solteira, SP - Brazil}

\author{M. B. Fontes}
\affiliation{Centro Brasileiro de Pesquisas F\'{\i}sicas, 22290-180, Rio de Janeiro, RJ, Brazil}

\author{E. Baggio-Saitovitch}
\affiliation{Centro Brasileiro de Pesquisas F\'{\i}sicas, 22290-180, Rio de Janeiro, RJ, Brazil}

\author{E. M. Bittar}
\affiliation{Centro Brasileiro de Pesquisas F\'{\i}sicas, 22290-180, Rio de Janeiro, RJ, Brazil}

\date{\today}

\begin{abstract}
The delicate balance between spin-orbit coupling, Coulomb repulsion and crystalline electric field interactions observed in Ir-based oxides is usually manifested as exotic magnetic behavior. Here we investigate the evolution of the exchange coupling between Co and Ir for partial La substitution by Ca in La$_{2}$CoIrO$_{6}$. A great advantage of the use of Ca$^{2+}$ as replacement for La$^{3+}$ is the similarity of its ionic radii. Thus, the observed magnetic changes can more easily be associated to electronic variations. A thorough investigation of the structural, electronic and magnetic properties of the La$_{2-x}$Ca$_{x}$CoIrO$_{6}$ system was carried out by means of synchrotron x-ray powder diffraction, muon spin rotation and relaxation ($\mu$SR), AC and DC magnetization, x-ray absorption spectroscopy (XAS), x-ray magnetic circular dichroism (XMCD), Raman spectroscopy, electrical resistivity and dielectric permittivity. Our XAS results show that up to 25\% of Ca substitution at the La site results in the emergence of Co$^{3+}$, possibly in high spin state, while the introduction of larger amount of Ca leads to the increase of Ir valence. The competing magnetic interactions resulting from the mixed valences lead to a coexistence of a magnetically ordered and an emerging spin glass (SG) state for the doped samples. Our $\mu$SR results indicate that for La$_{2}$CoIrO$_{6}$ a nearly constant fraction of a paramagnetic (PM) phase persists down to low temperature, possibly related to the presence of a small amount of Ir$^{3+}$ and to the anti-site disorder at Co/Ir sites. For the doped compounds the PM phase freezes below 30 K, but there is still some dynamics associated with the SG. The dielectric data obtained for the parent compound and the one with 25\% of Ca-doping indicate a possible magnetodielectric effect, which is discussed in terms of the electron hopping between the TM ions, the anti-site disorder at Co/Ir sites and the distorted crystalline structure.

\end{abstract}

%\pacs{75.47.Lx, 75.50.Lk, 75.30.Et, 75.30.-m}

\maketitle

\section{Introduction}

The recent increased interest in Ir-based oxides is ascribed to its extended $5d$ orbitals, giving rise to a unique interplay between the strong spin orbit coupling (SOC), the Coulomb repulsion and the crystal field splitting. In octahedral coordination, the increased SOC splits the $t_{2g}$ orbitals into a lower energy $j$ = 3/2 quartet and a higher energy $j$ = 1/2 doublet, leading to novel intriguing phenomena. For instance, the moderate electronic correlation effects open a gap within the $j$ = 1/2 doublet in Ir$^{4+}$ for Sr$_2$IrO$_{4}$, yielding a $j$ = 1/2 Mott insulating state \cite{Kim,Granado}, while for Cu$_2$IrO$_3$ it was found a Kitaev spin liquid ground state \cite{Abramchuk,Kenney}. In the case of Ir$^{5+}$, there is an open debate about the existence of excitonic magnetism in the anticipated nonmagnetic $j$ = 0 ground state of $5d^{4}$ Ir$^{5+}$ ions for Sr$_2$YIrO$_6$ and Ba$_2$YIrO$_6$ double perovskites (DP) \cite{Cao,Corredor,Fuchs,Brink}, argued to be consequence of an unusual interplay between strong noncubic crystal fields, local exchange interactions and SOC of intermediate strength. 

Concerning the DP systems, the combined action of Ir and other transition metal (TM) ions can also lead to interesting phenomena. In this context, the use of Co as the second TM is very attractive since the comparable energies associated to the crystal field splitting and to the interatomic exchange interactions can give rise to different valence and spin states in this ion. Apart from the high spin (HS) and low spin (LS) states, an intermediate spin (IS) state is also proposed for many Co-based perovskites. In addition, its substantial unquenched orbital magnetic moment can lead to strong spin-orbital-lattice coupling \cite{Raveau}. 

La$_2$CoIrO$_6$ is an example of material where the presence of Co and Ir can cause noteworthy physical properties. It is an insulating ferromagnetic (FM)-like system that exhibits a magnetodielectric effect at low temperatures \cite{Song}. Early reports of x-ray absorption spectroscopy (XAS), x-ray magnetic circular dichroism (XMCD) and neutron powder diffraction (NPD) proposed that its FM character would result from spin canting in the Co antiferromagnetic (AFM) sublattice, with the Ir AFM coupled to the Co ions \cite{Narayanan,Kolchinskaya}. However, recent XAS, XMCD and NPD studies have revealed AFM coupling between Co and Ir FM sublattices resulting in ferrimagnetic (FIM) behavior, and the manifestation of magnetoelastic effect associated to symmetric cooperative breathing distortions of the oxygen octahedra \cite{Lee,Noh}. In addition, the XAS investigation suggested that the Co 3$d$ $t_{2g}$ orbitals hybridize not only with the half filled $j_{eff}$ = 1/2 orbital of Ir$^{4+}$, but also with the fully empty $e_g$ orbitals, which would give rise to a persistent paramagnetism (PM) in La$_2$CoIrO$_6$. On the other hand, electronic changes induced by partial Ca$^{2+}$ to La$^{3+}$ substitution have led to a delicate balance between FM, AFM and spin glass (SG) phases in La$_{1.5}$Ca$_{0.5}$CoIrO$_{6}$, yielding up to three compensation temperatures and spontaneous exchange bias effect at low temperatures \cite{Coutrim}.

In a previous investigation of the structural and magnetic properties of La$_{2-x}$Ca$_{x}$CoIrO$_{6}$ compounds the anti-site disorder (ASD) at Co/Ir sites was shown to play a role on the magnetic coupling between these ions \cite{JSSC}. Here we produced samples with rather smaller ASD and focus our attention on the competition between SOC, the electronic correlation and the crystalline electric field interactions in Co- and Ir-based systems by investigating the effects of intermediate Ca$^{2+}$ to La$^{3+}$ substitution in La$_2$CoIrO$_6$, with the  parent compound being also investigated for comparison. A great advantage in the choice of Ca as the replacing ion is the fact its ionic radius is very close to that of La \cite{Shannon}. Thus the variations observed for different doping levels can more easily be associated to electronic changes. A thorough investigation of the structural, electronic and magnetic properties of the $x$ = 0, 0.5, 0.8 and 1.0 compounds was carried by means of synchrotron X-ray powder diffraction (SXRD), muon spin rotation and relaxation ($\mu$SR), Raman spectroscopy, ac and dc magnetization, XAS at Co-$K$ and Ir-$L_{3}$ edges, as well as XMCD at Ir-$L_{3}$ edge. Since the dielectric properties of DP compounds is usually related to the hopping of electrons between the TM ions, \textit{i.e.}, to the hybridization between the TM ions present in the system, we have also investigated the electrical resistivity and dielectric permittivity of the compounds.

Our results indicate that for the $x$ = 0 sample at least small fractions of Co$^{3+}$ and Ir$^{3+}$ ions are already present, and that up to 25\% of Ca$^{2+}$ substitution at La$^{3+}$ site acts mainly to increase the Co mean valence, whereas for $x$ = 0.8 and 1.0 compounds the effect of doping is to change the Ir formal valence. These electronic changes remarkably affect the magnetization in a way that the competing magnetic phases lead to SG behavior for the doped samples. The $\mu$SR results obtained for La$_2$CoIrO$_6$ parent compound indicate the persistence of a paramagnetic (PM) phase down to the lowest temperature investigated (5K), possibly associated to the presence of non-magnetic Ir$^{3+}$. On the other hand, for the Ca-doped samples there is the freezing of PM below 30 K, whilst one can still find some dynamics associated to the SG phase. All samples present insulating behavior, and the investigation of the dielectric properties for $x$ = 0 and 0.5 compounds indicated a possibile magnetodielectric effect. The electronic transport is discussed in terms of the electron hopping between the TM ions and of the distorted structure of these monoclinic materials.

\section{Experimental details}

The $x$ = 0, 0.5, 0.8 and 1.0 concentrations of the La$_{2-x}$Ca$_{x}$CoIrO$_{6}$ system were synthesized by conventional solid state reaction, as described in the Supplementary Material (SM) \cite{SM}. The SXRD data were recorded at the XPD beamline of the Brazilian Synchrotron Light Laboratory (LNLS) using a reflection geometry. The  XRD patterns were obtained at room temperature by a one-dimensional Mythen-1K detector (Dectris), using a wavelength $\lambda$ = 1.5498 $\textrm{\AA}$. The Rietveld refinements were performed using the program GSAS+EXPGUI \cite{GSAS}. The AC and DC magnetic measurements were carried out using a Quantum Design PPMS-VSM magnetometer. 

Muon spin rotation and relaxation ($\mu$SR) experiments were performed at the Swiss Muon Source of Paul Scherrer Institut, Switzerland, using the nearly 100$\%$ spin-polarized positive muon beam at the GPS instrument. Spectra were measured on polycrystalline pieces in zero field (ZF) and weak transverse field (wTF, field applied perpendicular to the initial muon spin direction) modes. Samples were wrapped in aluminized mylar foils and mounted within the beam spot between the legs of an ultrapure copper fork. For an efficient stopping of muons in the sample a 250 $\mu$m ultrapure aluminum foil of proper size in front of the sample served as degrader. Thermal contact was achieved by helium exchange gas. The data were acquired at several temperatures between 5 and 300 K. For the wTF experiments a field $H$ = 50 Oe was applied.

XAS measurements at the Co-$K$ and Ir-$L_{3}$ edges, as well as XMCD at the Ir-$L_{3}$ edge, were performed in the dispersive X-ray absorption (DXAS) beamline at LNLS \cite{DXAS}. The XAS and XMCD spectra were investigated with the samples in the form of powder, but prior to the powdering of the bulk the pellets' surfaces were scraped in order to avoid the influence of its redox in the results. The circular polarization rate for the XMCD measurements was of approximately 80\%. Unpolarized Raman scattering measurements were performed at low temperature on a Jobin Yvon T64000 triple 1800 mm$^{-1}$ grating spectrometer equipped with a liquid N$_{2}$-cooled multichannel CCD detector. The excitation was achieved with a 488 nm Ar$^{+}$ laser line in a quasi-backscattering configuration. 

Electrical resistivity was measured at the PPMS low-frequency ac resistance bridge in a four-contact configuration, using platinum wires and silver paste for the contacts. An Agilent 4284A LCR meter was used to electrical measurements of complex admittance ($Y = G + jB$), defined in terms of the conductance ($G$) and admittance ($B$), to obtain the real ($\varepsilon'$) and imaginary part ($\varepsilon''$) of dielectric permittivity ($\varepsilon = \varepsilon' + \varepsilon''$), the imaginary part ($Z$'') of complex impedance ($Z^{*}= Z'-jZ''$) and the ac electric conductivity $\sigma(\omega)$ at different frequencies (10$^{2}$ - 10$^{6}$ Hz) and temperatures (27 K - 300 K).

\section{Results and discussions}

\subsection{Synchrotron x-ray diffraction}

\begin{figure}
\begin{center}
\includegraphics[width=0.47 \textwidth]{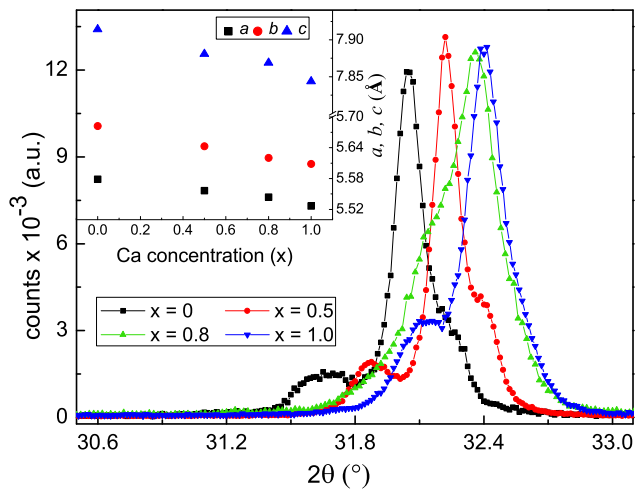}
\end{center}
\caption{Magnified view of the (020),(112),(11\={2}) and (200) Bragg reflections in the SXRD patterns of La$_{2-x}$Ca$_{x}$CoIrO$_{6}$. The inset shows the $a$, $b$ and $c$ lattice parameters as a function of Ca concentration, $x$.}
\label{Fig_SXRD}
\end{figure}

The SXRD patterns and the Rietveld analysis confirmed the formation of single phase DP samples for $x$ = 0, 0.5, 0.8 and 1.0 concentrations of La$_{2-x}$Ca$_{x}$CoIrO$_{6}$ (herein described by their Ca-concentrations, $x$), as it can be seen in SM \cite{SM}. All compounds belong to the monoclinic $P2_{1}/n$ space group, as expected since the Ca$^{2+}$ and La$^{3+}$ ionic radii are very close in XII coordination (1.34 and 1.36 \AA, respectively \cite{Shannon}). Fig. \ref{Fig_SXRD} shows the shift of 020, 200, 112 and 11$\overline{2}$ Bragg reflections toward higher angles as the Ca concentration increases, resulting from the shrinkage of the unit cell as shown in the inset. Although the Ca$^{2+}$ ionic radius is slightly smaller than that of La$^{3+}$, such shrinkage is most probably related to the systematic increase in the oxidation state of the TM ions, for which the changes in the ionic radii are expected to be larger \cite{Shannon}.

The octahedral tilts resulting from the shrinkage of the lattice parameters lead to the expected decrease of the Co--O--Ir bond angle, as can be seen in Table \ref{T_SXRD}. This table also shows the ASD at Co/Ir sites. The $\sim$4.4\% value obtained for the $x$ = 0 sample is nearly the half of the value found for a sample produced at a rather different furnace temperature \cite{JSSC}, suggesting that this system is very sensitive to the synthesis route, as usual in DP compounds \cite{Raveau,Vasala}. For the doped samples, it can be seen an initial increase of ASD for $x$ = 0.5, followed by a tendency of decrease for larger Ca concentrations. As will be discussed next, this is in agreement with the XAS and magnetization results, which indicated the increase of Co average valence for $x$ = 0.5 sample in comparison to the parent compound. This makes the Co and Ir ions chemically more similar, increasing the chance of permutation between these ions. On the other hand, for larger Ca doping the XAS show a tendency for an increasing  average valence of Ir, making the Co/Ir permutation less likely. 

\begin{table}
\renewcommand{\arraystretch}{1.2}
\caption{Main results obtained from the Rietveld refinements of the SXRD data.}
\label{T_SXRD}
\resizebox{\columnwidth}{!}{
\begin{tabular}{ccccc}
\hline \hline
Ca concentration ($x$): & 0 & 0.5 & 0.8 & 1.0 \\
\hline
$a$ (\AA) & 5.5786(3) & 5.5566(1) & 5.5438(6) & 5.5269(2) \\

$b$ (\AA) & 5.6821(3) & 5.6427(2) & 5.6202(6) & 5.6086(3) \\

$c$ (\AA) & 7.9145(4) & 7.8808(2) & 7.8688(9) & 7.8434(4) \\

$\beta$ ($^{\circ}$) & 89.96(1) & 89.98(1) & 89.76(1) & 89.94(1) \\

$V$ (\AA$^{3}$) & 250.87(3) & 247.09(2) & 245.17(5) & 243.13(2) \\

ASD (\%) & 4.4(3) & 7.3(3) & 6.2(6) & 5.6(4) \\

$<$Co-O-Ir$>$ ($^{\circ}$) & 154.6(1) & 149.4(1) & 148.6(2) & 146.5(6) \\

$R_p$ (\%) & 12.7 & 8.5 & 11.3 & 9.4 \\

$\chi^{2}$ & 1.8 & 2.3 & 2.0 & 2.2 \\

\hline \hline
\end{tabular}}
\end{table}

\subsection{X-ray absorption spectroscopy}

\begin{figure}
\begin{center}
\includegraphics[width=0.45 \textwidth]{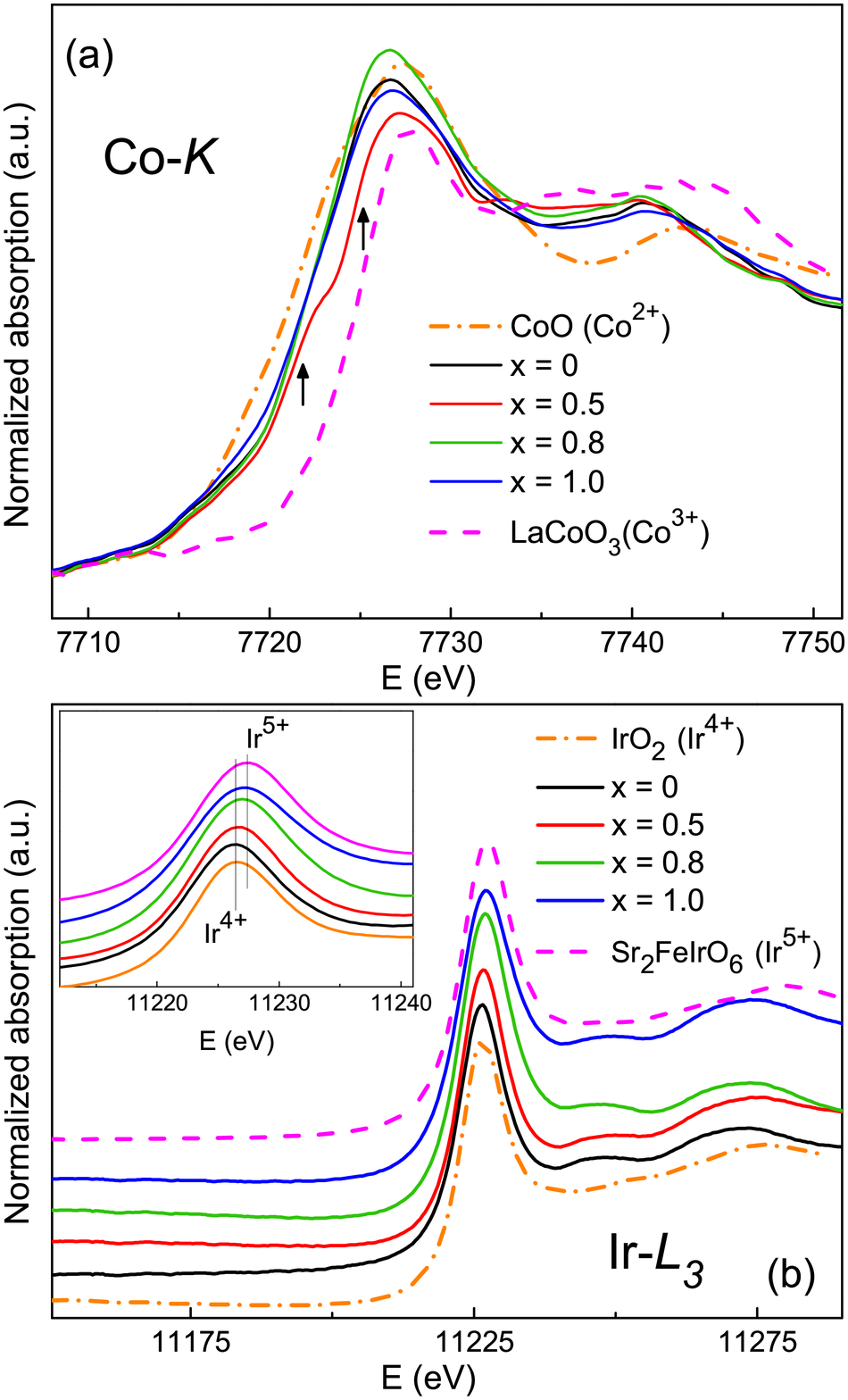}
\end{center}
\caption{(a) Co-$K$ edge XAS spectra of La$_{2-x}$Ca$_{x}$CoIrO$_{6}$ samples. The spectra of CoO and LaCoO$_3$ are also displayed as reference for Co$^{2+}$ and Co$^{3+}$, respectively. The presence of two white lines, indicating Co$^{2+}$/Co${3+}$ mixed valence state, is highlighted by arrows in the curve of $x$ = 0.5 sample. (b) Ir-$L_3$ spectra, together with those of IrO$_2$ and Sr$_{2}$FeIrO$_{6}$, reference samples for Ir$^{4+}$ and Ir$^{5+}$, respectively. The inset shows a magnified view of the absorption edges.}
\label{Fig_XAS}
\end{figure}

Fig. \ref{Fig_XAS}(a) shows the normalized Co-$K$ edge XAS of all investigated samples, together with the spectra of CoO and LaCoO$_{3}$ used as reference samples for Co$^{2+}$ and Co$^{3+}$, respectively. It can be seen that the spectral shape of $x$ = 0 is similar to that of CoO reference sample, indicating a Co$^{2+}$ configuration for the majority of ions. However, it can be noticed that the spectra's white line lies in between those of CoO and LaCoO$_{3}$. Although some differences between the absorption curves of La$_{2}$CoIrO$_{6}$ and CoO should be expected since they present distinct structures, a $\sim$0.6 eV shift of the white line's position indicates that at least a small fraction of Co$^{3+}$ is present already for the parent compound. The 25\% of Ca substitution at the La site leads to the increase of this fraction, as can be clearly observed by the presence of two distinct white lines, marked by arrows in Fig. \ref{Fig_XAS}(a). The curves of $x$ = 0.8 and 1.0 samples are similar to that of $x$ = 0, indicating a tendency of return to Co$^{2+}$ configuration. Since the insertion of Ca represents a hole doping, this tendency toward Co$^{2+}$ should signify an increase of Ir valence to ensure the charge balance in La$_{2-x}$Ca$_{x}$CoIrO$_{6}$.

In order to check whether the variations in the Co valence induced by Ca to La substitution are accompanied by changes in the Ir electronic state, we performed XAS at Ir-$L_3$ edge. Fig. \ref{Fig_XAS}(b) shows the normalized spectra of all samples of interest, and also those of IrO$_{2}$ and Sr$_{2}$FeIrO$_{6}$, used as reference samples for Ir$^{4+}$ and Ir$^{5+}$, respectively. It can be noticed a great similarity between the curves, which is a characteristic feature of the Ir-$L_3$ edge, being related to the fact the 5$d$ transition metals have more diffuse valence orbitals compared to 3$d$ ones \cite{Kolchinskaya,Liu}. However, a magnified view of the absorption edge depicted in the inset of \ref{Fig_XAS}(b) shows some subtle differences between the curves. The position of the absorption edge for $x$ = 0 sample is slightly shifted to the left with respect to that of IrO$_{2}$ sample, suggesting that a small fraction of Ir$^{3+}$ may be present. This was expected since the Co-$K$ edge XAS indicated the presence of Co$^{3+}$, albeit the oxygen stoichiometry was not checked. The spectra of $x$ = 0.5 is slightly shifted to higher energies, whereas for $x$ = 0.8 and 1.0 the displacements are more pronounced, indicating the tendency toward Ir$^{5+}$ for these samples.

As aforementioned, structural changes can usually lead to variations in the shape of the absorption spectra. However, such large shifts as those observed for the Co-$K$ edge are most likely related to valence changes. In order to get quantitative estimates of the Co formal valences, we compared the white line position in energy for the samples (see SM \cite{SM}). Assuming a linear variation from Co$^{2+}$ to Co$^{3+}$, we obtain approximately 2.1+, 2.3+, 2.2+ and 2.1+ for $x$ = 0, 0.5, 0.8 and 1.0 respectively. A similar procedure for Ir-$L_3$ edge resulted in approximately 3.8+, 4.1+, 4.5+ and 4.7+ for $x$ = 0, 0.5, 0.8 and 1.0 respectively. These values would lead to oxygen vacancies in the La$_{2-x}$Sr$_{x}$CoIrO$_{6-\delta}$ samples, with $\delta$ ranging from 0.05 to 0.1.  Notwithstanding the inaccuracy of these rough estimates and the lack of a direct investigation of the oxygen content, non-stoichiometry in the oxygen site is usually observed in DPs \cite{Vasala}, being thus a plausible scenario. Further investigation using other techniques as X-ray photoelectron spectroscopy and thermogravimetry are necessary to confirm these results.

\begin{figure}
\begin{center}
\includegraphics[width=0.47 \textwidth]{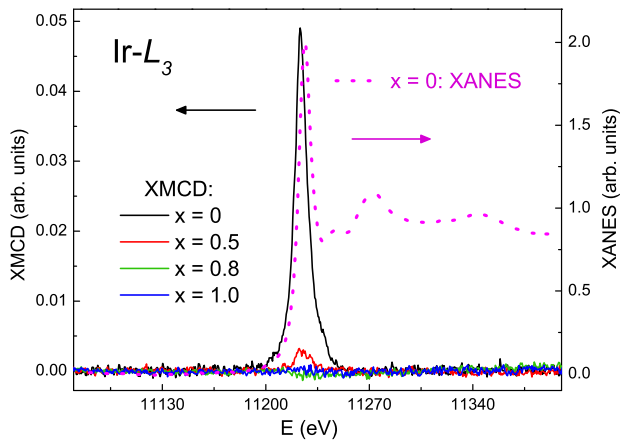}
\end{center}
\caption{60 K Ir-$L_3$ XANES spectrum for $x$ = 0 and XMCD spectra for $x$ = 0, 0.5, 0.8 and 1.0 samples.}
\label{Fig_XMCD}
\end{figure}

The tendency toward Ir$^{5+}$ with increasing Ca-content could be confirmed by the XMCD curves, carried for Ir-$L_3$ edge with an applied magnetic field $H$ = 9 kOe at $T$ = 60 K. The XMCD signals of all investigated samples are shown in Fig. \ref{Fig_XMCD}, together with the XANES spectrum of $x$ = 0. The signal, associated to the Ir$^{4+}$ $j$ = 1/2 state for the parent compound, is about 40 times smaller than the XANES absorption signal. Its negative sign signifies a negative moment, \textit{i.e.}, the Ir$^{4+}$ moment is in opposite direction with respect to the applied field. For the $x$ = 0.5 sample one can see a great reduction in the XMCD signal. This is in contrast with the XAS results which have shown that the Ca to La substitution affects mainly the Co valence state. Thus, such weaken in the XMCD signal is indicative of changes in the nature of the magnetic coupling between the TM ions in this system. N. Narayanan \textit{et al.} by means of NPD have proposed that Co and Ir form two AFM sublattices in La$_{2-x}$Sr$_{x}$CoIrO$_{6}$ doped samples \cite{Narayanan}, whereas L. Coutrim \textit{et al.} suggested that the small Ir$^{4+}$ signal observed for La$_{1.5}$Ca$_{0.5}$CoIrO$_{6}$ comes from frustration of Ir moment due to the AFM coupling of its Co neighbors \cite{Coutrim}. For $x$ = 0.8 and 1.0 samples the XMCD signal is negligible, in accordance with the tendency toward non-magnetic Ir$^{5+}$ and also with the conjecture of AFM at Ir sublattice.

\subsection{Raman spectroscopy}

\begin{figure}
\begin{center}
\includegraphics[width=0.47 \textwidth]{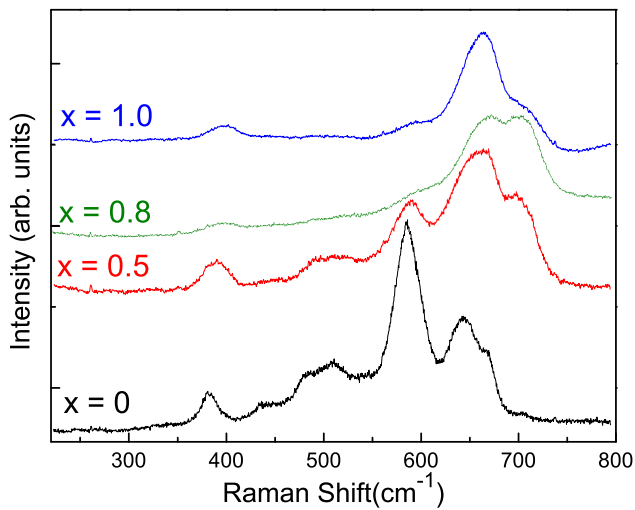}
\end{center}
\caption{Raman spectra of the La$_{2-x}$Ca$_{x}$CoIrO$_{6}$ samples at 23 K.}
\label{Fig_Raman}
\end{figure}

Since the properties of interest for La$_{2-x}$Ca$_{x}$CoIrO$_{6}$ are observed below the magnetic ordering transitions, we investigated the Raman spectroscopy at low temperatures. Fig. \ref{Fig_Raman} shows the Raman spectra at 23 K, where it can be noticed a number of peaks associated to the La$_{2-x}$Ca$_{x}$CoIrO$_{6}$ system. Comparing the spectra with those of similar DP compounds as Sr$_{2}$CoIrO$_{6}$, Ba$_{2}$YIrO$_{6}$, La$_{2}$CoMnO$_{6}$, Sr$_{2}$CrReO$_{6}$ and Ba$_{2}$FeReO$_{6}$ \cite{Esser,Singh,Iliev,Fournier,PRB2019,Moreira} one can associate the peaks at $\sim$380, 440, 490 and 510 cm$^{-1}$ for $x$ = 0 to bending modes and those at $\sim$650 and 665 cm$^{-1}$ to stretching modes. The peak at $\sim$590 cm$^{-1}$ may be associated to either anti-stretching or bending mode vibrations of (Co/Ir)O$_6$ octahedra, while that at $\sim$700 cm$^{-1}$ may be related to a breathing mode of the (Co/Ir)O$_6$ octahedra. 

An interesting result obtained from the Raman data is that the $\sim$490 and 510 cm$^{-1}$ peaks appearing for the $x$ = 0 sample weaken progressively as the Ca-doping increases, becoming negligible for the $x$ = 1.0 compound. These modes are observed in the CaIrO$_3$ Ir$^{4+}$ system \cite{Hustoft} but not in the Ba$_{2}$YIrO$_{6}$ Ir$^{5+}$ one \cite{Singh}. Conversely, the opposite trend was found for the $\sim$700 cm$^{-1}$ peak, where this is negligible for $x$ = 0 and becomes intense for the Ca-doped samples. The frequency of this mode is a marker of the Co--O and Ir--O spring constants for stretching vibrations, which tend to become harder for larger cationic values such as Ir$^{5+}$ and Co$^{3+}$ with respect to Ir$^{4+}$ and Co$^{2+}$. Indeed, oxides presenting Ir$^{5+}$ and B$^{3+}$ as Sr$_{2}$CoIrO$_{6}$ and Ba$_{2}$YIrO$_{6}$ may exhibit stretching vibrations in the spectral region $\sim$700 cm$^{-1}$ and beyond \cite{Esser,Singh}, contrary to Ir$^{4+}$ and  Co$^{2+}$ oxides that have not been reported to present phonons at such high energies \cite{Hustoft,Murthy3,Bao}, being thus this mode most probably related to the frequency of the oxygen spring stretching vibration in the Co$^{3+}$--O--Ir$^{5+}$ coupling. Therefore, the Raman data are in agreement with the increase of Co$^{3+}$ and/or Ir$^{5+}$ portions with Ca-doping indicated by the XAS measurements. This is also observed in the AC and DC magnetization measurements, as will be discussed next. 

\subsection{AC and DC magnetization}

\begin{figure}
\begin{center}
\includegraphics[width=0.465 \textwidth]{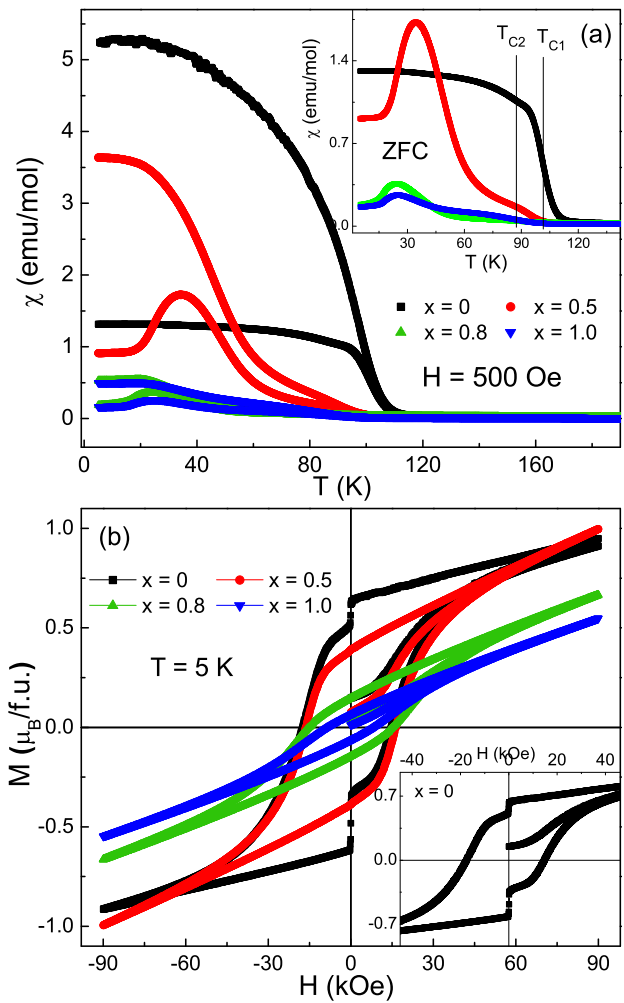}
\end{center}
\caption{(a) ZFC and FC $M(T)$ curves for La$_{2-x}$Ca$_{x}$CoIrO$_{6}$, measured at $H$ = 500 Oe. The inset shows a magnified view of the ZFC curves, evidencing $T_{C1}$, $T_{C2}$ and $T_f$. (b) $M(H)$ curves measured at $T$ = 5 K. The inset shows a magnified view of the $x$ = 0 curve near the $H$ = 0 region.}
\label{Fig_PPMS}
\end{figure}

The magnetization measurements as a function of temperature [$M(T)$] were recorded in both zero field cooled (ZFC) and field cooled (FC) modes, with $T$ ranging from 5 to 400 K at $H$ = 500 Oe. The results displayed in Fig. \ref{Fig_PPMS}(a) reveal magnetic order below 100 K for all compositions, with the magnetization being systematically reduced with increasing Ca content. The differences between the ZFC and FC curves indicate the presence of FM components, as confirmed by the magnetization as a function of $H$ [$M(H)$] curves, Fig. \ref{Fig_PPMS}(b). 

For $x$ = 0, the FM-like character is attributed to the antiparallel orientations of Co$^{2+}$ and Ir$^{4+}$ FM sublattices, resulting in FIM behavior \cite{Lee,Noh}. For the $x$ = 0.5 sample, although the increased mixed valence state weakens the magnetization, there is still a fraction of Co$^{2+}$--Ir$^{4+}$ coupling leading to FIM. The cases of $x$ = 0.8 and 1.0 compounds are similar to that of La$_{2}$CoPtO$_{6}$(note that both Ir$^{5+}$ and Pt$^{4+}$ are expected to have $j_{eff}$ = 0) \cite{Noh}. According to the mechanism of orbital hybridization, the virtual electron hopping from Co 3$d$ $t_{2g}$ to Ir 5$d$ $j_{eff}$ = 0 is not spin-selective, leading to the reduced magnetization observed in Fig. \ref{Fig_PPMS}. For the La$_{2-x}$Sr$_{x}$CoIrO$_{6}$ resemblant system the weak magnetization of the Sr-doped compounds was attributed to canted AFM \cite{Narayanan}.

A close inspection of the ZFC curves, inset of Fig. \ref{Fig_PPMS}(a), revealed the appearance of two anomalies, $T_{C1}$ and $T_{C2}$, at the $\sim$100 K region. This becomes evident already for the $x$ = 0.0 sample, while for the doped compounds the proximity between the anomalies prevent a trustworthy estimate of the transition temperatures from the DC $M(T)$ data carried with $H$ = 500 Oe. Reliable determination of $T_{C1}$ and $T_{C2}$ were obtained from the inflection points of the AC susceptibility curves, as will be discussed next. The presence of two nearby anomalies can be interpreted as a rough independency between the ordering temperatures of Co and Ir sub-lattices. Recent report of first principle calculation have suggested that the Ir-$d$ states are hardly affected by the Co-$d$ ones \cite{Ganguly}. Since the XAS and SXRD results indicated mixed valence states and ASD for all investigated samples, there is also the possibility of $T_{C2}$ being related to the magnetic transition temperature of other nearest neighbor couplings such as Co$^{2+}$--Co$^{2+}$, Co$^{3+}$--Co$^{3+}$, Ir$^{4+}$--Ir$^{4+}$ or Co$^{2+}$--Co$^{3+}$. Although these couplings are predicted to be AFM by the Goodenough-Kanamory-Anderson rules \cite{GKA}, the last one would give rise to FIM behavior.

For the $x$ = 0 compound, the fit of the PM region to the Curie-Weiss (CW) law yielded a CW temperature $\theta_{CW}$ = -17 K (see SM \cite{SM}), the small negative $\theta_{CW}$ being in accordance with FIM behavior. For the Ca-doped samples the largely negative  $\theta_{CW}$ values indicate a trend to strong AFM coupling with increasing Ca content (see Table \ref{T_mag}). The values observed in Table \ref{T_mag} are somewhat different to those found for La$_{2-x}$Ca$_{x}$CoIrO$_{6}$ samples prepared by a distinct synthesis route \cite{JSSC}. Although sample dependence is an usual feature of DPs \cite{Vasala}, in special Co-based ones \cite{Raveau}, these discrepancies are most likely related to the fact the $M(T)$ curves reported in Ref. \onlinecite{JSSC} were investigated up to 300 K while in this work it goes up to 400 K. 

The effective magnetic moment $\mu_{eff}$ = 4.6 $\mu_{B}$/f.u obtained from the CW fit for the parent compound is in agreement with previous reports \cite{Song,Narayanan}. The formal Co and Ir valences estimated from the XAS results can be applied in the usual equation for systems with two or more different magnetic ions, $\mu$ = $\sqrt{{\mu_1}^2 + {\mu_2}^2 + ...}$, to evaluate the expected magnetic moment. Applying $\mu_{Co^{2+}}$ = 3.9 $\mu_B$ and $\mu_{HSCo^{3+}}$ = 4.9 $\mu_B$ for the spin-only (SO) approximation and  $\mu_{Ir^{4+}}$ = 1.7 $\mu_B$ yields  $\mu$ = 4.0 $\mu_{B}$/f.u. if LS configuration is assumed for Co$^{3+}$ and $\mu$ = 4.3 $\mu_{B}$/f.u. if HS Co$^{3+}$ is considered. Both values are rather below the experimental result. Previous studies in La$_{2}$CoIrO$_{6}$ indicated deviations from the SO values due to a non-negligible orbital contribution to the Co moment \cite{Song,Kolchinskaya,JSSC}. Using the standard magnetic moments of Co$^{2+}$ (4.8 $\mu_B$) and HS Co$^{3+}$ (5.4 $\mu_B$) \cite{Ashcroft} one obtain  $\mu$ = 5.1 $\mu_{B}$/f.u. for Co$^{3+}$ in HS configuration and $\mu$ = 4.8 $\mu_{B}$/f.u. for LS Co$^{3+}$, closer to the experiment. 

The great increase of $\mu_{eff}$ observed in Table \ref{T_mag} for $x$ = 0.5 can not be accounted for the Ir valence changes, being most likely related to the emergence of HS Co$^{3+}$. The expected moment calculated in the SO approximation lies far below the experimental value, as well as when LS Co$^{3+}$ is considered. Using the standard Co$^{2+}$ and HS Co$^{3+}$ moments yields  $\mu$ = 5.2 $\mu_{B}$/f.u., fairly close the experimental $\mu_{eff}$ = 5.4 $\mu_{B}$/f.u.. For the $x$ = 0.8 and 1.0 samples the expected moments calculated within the SO approximation are also much lower than the experiment, as do those calculated assuming LS Co$^{3+}$. Using the standard Co$^{2+}$ and HS Co$^{3+}$ moments results in $\mu$ = 5.1 and 5.0 $\mu_{B}$/f.u. for $x$ = 0.8 and 1.0, much closer to the experimental values. Despite the subtle discrepancies observed between the observed and calculated moments, both the experimental and theoretical values follow the same trend with increasing Ca-content, as can be seen in Table \ref{T_mag}. This endorses the evolution of Co and Ir valences found from XAS.

The inset of Fig. \ref{Fig_PPMS}(a) highlights the emergence of an anomaly in the ZFC curve of $x$ = 0.5 sample at lower temperatures. This is also present, to a lesser extent, in the curves of $x$ = 0.8 and 1.0 samples. This peak was thoroughly investigated in Ref. \onlinecite{Coutrim}, where it was properly associated to the emergence of SG state.  The AC susceptibility and $\mu$SR results will confirm that glassy magnetism is also present in the $x$ = 0.8 and 1.0 compounds.  

\begin{table}
\renewcommand{\arraystretch}{1.2}
\caption{Main results obtained from $M(T)$, $M(H)$ and AC susceptibility curves. The $\chi_{ac}$ results for $x$ = 0.5 sample were extracted from Ref. \onlinecite{Coutrim}.}
\label{T_mag}
\resizebox{\columnwidth}{!}{
\begin{tabular}{c|cccc}
\hline \hline
Ca concentration (x) & 0 & 0.5 & 0.8 & 1.0 \\
\hline
$T_{C1}$ (K) & 98 & 96 & 95 & 96 \\

$T_{C2}$ (K) & 87 & 86 & 81 & 77 \\

$\theta_{CW}$ (K) & -17 & -53 & -75 & -85 \\

$\mu_{eff}$ ($\mu_B$/f.u.) & 4.6 & 5.4 & 5.3 & 5.1 \\

\hline

$M(90 kOe)$ ($\mu_B$/f.u.) & 0.95 & 1.00 & 0.67 & 0.55 \\

$M_r$ ($\mu_B$/f.u.) & 0.62 & 0.39 & 0.15 & 0.06 \\

slope (of linear fit) & 3.2$\times$10$^{-6}$ & 6.1$\times$10$^{-6}$ & 5.1$\times$10$^{-6}$ & 4.7$\times$10$^{-6}$ \\

$M_s$ ($\mu_B$/f.u.) & 0.66 & 0.44 & 0.21 & 0.12 \\

$H_C$ (Oe) & 17554 & 16437 & 15191 & 8478 \\

\hline

$T_{sg}$ (K) & - & 26.6 & 25.7 & 25.9 \\

$\tau_0$ (s) & - & 1.5$\times$10$^{-13}$ & 7.5$\times$10$^{-14}$ & 8.0$\times$10$^{-14}$ \\

\textit{z}$\nu$ & - & 6.5 & 5.8 & 6.5 \\

\hline \hline
\end{tabular}}
\end{table}

The $M(H)$ loops, carried at 5 K with a maximum applied field $H_{m}$ = 90 kOe, are diplayed in Fig. \ref{Fig_PPMS}(b). The curve of $x$ = 0 sample shows a non-conventional behavior, with abrupt changes in the magnetization near the $H$ = 0 regions. This might be related to the presence of two AFM coupled sublattices, as suggested by the XMCD and $M(T)$ curves. For the Ca-doped samples, the decreased concentration of Ir$^{4+}$ and/or Co$^{2+}$, together with the inhomogeneity naturally enhanced by the inclusion of a new ion, may be preventing the observation of such effect. 

It can be noticed the lack of a complete magnetic saturation for all $M(H)$ curves. From the extrapolation of the linear fit of the high field regions of the curves ($H$ $>$ 80 kOe), the FM component of each sample can be estimated ($M_{s}$). The results, together with the slope of the fitted lines, the $M$ value at $H$ = 90 kOe [$M(90 kOe)$], the remanent magnetization ($M_{r}$) and the coercive fields ($H_{C}$), are displayed in Table \ref{T_mag}. The small values of $M_{s}$, $M(90 kOe)$ and $M_{r}$ reinforce the strong FIM character of the system. The slightly larger $M(90 kOe)$ value of $x$ = 0.5 in comparison to $x$ = 0 is in agreement with its larger $\mu_{eff}$, as well as the decrease of the magnetization for the $x$ = 0.8 and 1.0 samples.

An interesting result observed in Table \ref{T_mag} is the decrease of $H_{C}$ with increasing Ca-content, \textit{i.e.}, $H_{C}$ decreases with the reduction of the lattice parameters. Previous reports have observed the increase of $H_{C}$ with the decrease of the A-site ionic radius in other DPs, which was interpreted in terms of the enhancement of the orbital contribution to the magnetic moment of the TM ion \cite{Sikora,PRB2019}. In our case, the opposite behavior was found. The decrease of $H_C$ from $x$ = 0 to 0.5 might be related to the increase of HS Co$^{3+}$ fraction, whereas for $x$ = 0.8 and 1.0 the changes are most probably associated to the decreased portion of Ir$^{4+}$. A recent investigation of hydrostatic pressure applied in DP compounds has shown variations of $H_{C}$ related to pressure-induced changes in the crystalline field \cite{Haskel}. In the case of La$_{2-x}$Ca$_{x}$CoIrO$_{6}$, this scenario is plausible since the Ca to La substitution is a chemical pressure that induce changes in the structure and in the Co and/or Ir valences, which in turn will certainly affect the crystal field.

\begin{figure}
\begin{center}
\includegraphics[width=0.465 \textwidth]{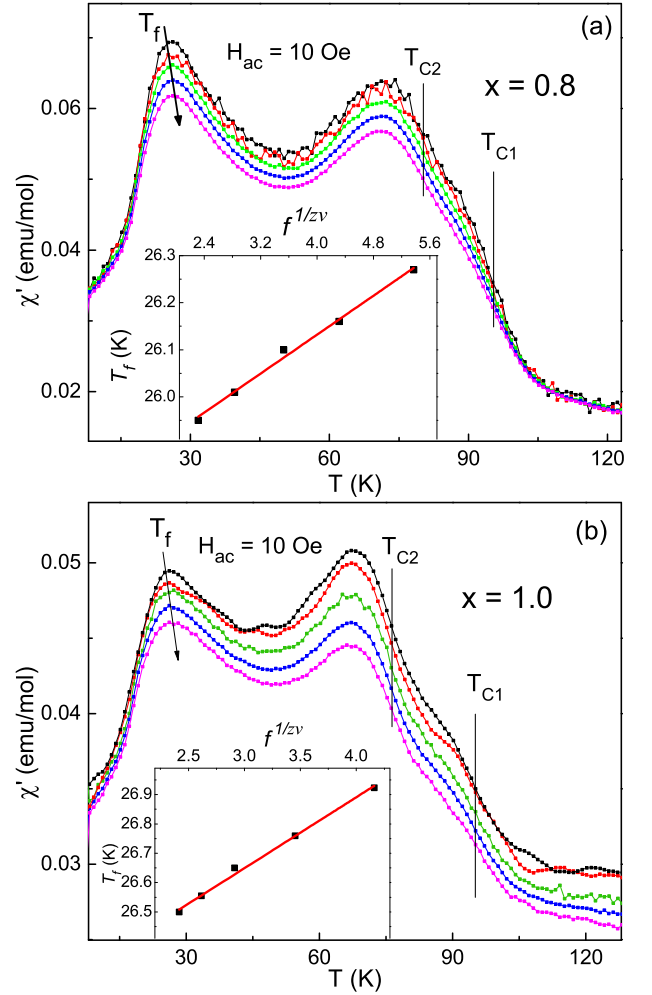}
\end{center}
\caption{$\chi'$ as a function of $T$ for (a) $x$ = 0.8 and (b) $x$ = 1.0 samples. The vertical lines are guidelines for $T_{C1}$ and $T_{C2}$ determined from the inflection points in the curves. The insets show $T_f$ as a function of $f$, where the solid lines represent the best fits to the data using Eq. \ref{Eq1}.}
\label{Fig_chiAC}
\end{figure}

The low-$T$ anomaly observed in the ZFC $M(T)$ curve of $x$ = 0.5 concentration was thoroughly investigated in Ref. \onlinecite{Coutrim}, where this peak was associated to the appearance of SG state caused by the competing magnetic phases and ASD. In order to verify the presence of SG-like phase in the other samples, we performed AC magnetic susceptibility ($\chi_{ac}$) at $H_{ac}$ = 10 Oe, with several frequencies ($f$) ranging from 100 to 10000 Hz. For $x$ = 0 it was not observed any indicative of SG behavior, while for the $x$ = 0.8 and 1.0 compounds non-negligible changes in the curves could be noticed. Fig. \ref{Fig_chiAC} displays the real part of the susceptibilities ($\chi$') for $x$ = 0.8 and 1.0 compounds, where the intensities of the low-$T$ anomalies decrease and the peak positions ($T_f$) shift to higher $T$ with increasing $f$. This is a signature of glassy magnetic behavior \cite{Mydosh}. The intensity of the high-$T$ anomalies also decrease with increasing $f$, but it was not observed any measurable shift of the peak positions, being thus these peaks associated to the ordinary magnetic transitions discussed above\cite{Fujiki,Balanda}. 

The $T_f$ as a function of $f$ curves of each sample could be well fitted by the power law equation of the dynamic scaling theory, commonly used to investigate SG-like systems \cite{Mydosh,Souletie}
\begin{equation}
\frac{\tau}{\tau_{0}}=\left[\dfrac{(T_{f} - T_{g})}{T_{g}}\right]^{-z\nu} \label{Eq1}
\end{equation}
where $\tau$ is the relaxation time corresponding to the measured frequency, $\tau_{0}$ is the characteristic relaxation time of spin flip, $T_{g}$ is the SG-like transition temperature ($T_f$ as $f$ tends to zero), $z$ is the dynamical critical exponent and $\nu$ is the critical exponent of the correlation length. The main results obtained from the fittings are displayed in table \ref{T_mag}. The $\tau_{0}$ and $z\nu$ values are typically found for SG systems \cite{Souletie,Malinowski}. This confirms that for the Ca-doped compounds there is the emergence of a SG phase below the conventional ordering temperatures.

\subsection{Muon spin rotation and relaxation}

In order to further clarify the complex magnetic behavior of La$_{2-x}$Ca$_{x}$CoIrO$_{6}$ samples, we performed $\mu$SR experiments using wTF and ZF modes. The experiments give the time dependent asymmetries $A(t)$ of the decay positron count rates from which we can derive the time dependent polarization of muons after implantation at time $t$=0:
\begin{equation}
A(t)/A(0) = G_{z}(t).
\label{Eq2}
\end{equation}

\begin{figure}
\begin{center}
\includegraphics[width=0.46 \textwidth]{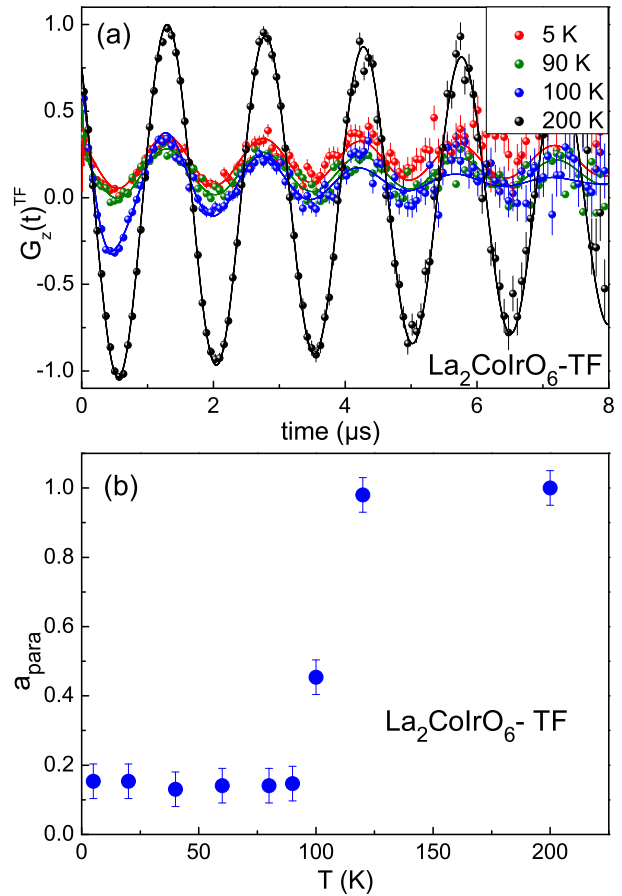}
\end{center}
\caption{(a) $\mu$SR rotation patterns of La$_{2}$CoIrO$_{6}$ in a weak applied transverse field (wTF) of 50 Oe at some selected temperatures; (b) variation of weakly damped PM fraction with temperature derived from wTF spectra.}
\label{Fig_muSRTF0}
\end{figure}

Fig. \ref{Fig_muSRTF0}(a) below shows the wTF spectra of $x$ = 0 at some selected temperatures (measured in a transverse field $H_{TF}$ = 50 Oe). Comparing the spectrum at 200 K with those at 100 K and below, one can note a decrease in the initial asymmetry and an increase of damping of the rotation signal, confirming the onset of magnetic ordering at $\sim$100 K, as observed in the $M(T)$ curves. The wTF signal can be fitted with the function
\begin{equation}
G_{z}(t)^{TF} = a_{para} \cdot G_{z}(t)^{TF}_{para} + a_{fast} \cdot e^{-\lambda_{fast}} + a^{TF}_{BG},
\label{Eq3}
\end{equation}
where the first term corresponds to the damped oscillating part due to muons in paramagnetic surrounding, $G_{z}(t)^{TF}_{para}$ = $e^{-(\lambda_{para}t)}cos(2\pi\nu_{\mu}t$), with the frequency $\nu_{\mu}$ = $\gamma_{\mu}B_{\mu}$/2$\pi$. $\lambda_{para}$ is the damping parameter, $B_{\mu}$ is the field acting at the muon site, \textit{i.e.} the field applied perpendicular to the muons initial polarization slightly modified by Knight shift, $\gamma_{\mu}$ = 135.54 $\mu$s$^{-1}$/T is the muon gyromagnetic ratio. 

The second term in Eq. \ref{Eq3} represents the fast damping in the initial part of the spectra leading to the apparent loss of the asymmetry of the rotating signal. It is caused by a wide distribution of strong local fields from magnetically ordered electronic moments with respect to the small applied field. $a_{para}$ and $a_{fast}$ are the partial asymmetries representing PM and two third of the long-range ordered volume fractions. One third of the long-range ordered fraction (the ``1/3 tail'') is comprised in an undamped background $a_{BG}$, seen from the upward shift of the rotation signals in Fig. \ref{Fig_muSRTF0}(a). Fig. \ref{Fig_muSRTF0}(b) shows the fast decrease of $a_{para}$ around 100 K. However, it can be noticed that a fraction of the rotation signal from the PM phase persists down to the lowest measured temperature, 5 K. Part of it is due to an undamped signal from the sample holder and a small contribution from muons stopped in the degrader foil (though this should be negligible for the chosen thickness). The major part of this oscillating contribution is atrributed to a magnetically not ordered fraction that is also found in ZF measurements as will be discussed below.

\begin{figure}
\begin{center}
\includegraphics[width=0.5 \textwidth]{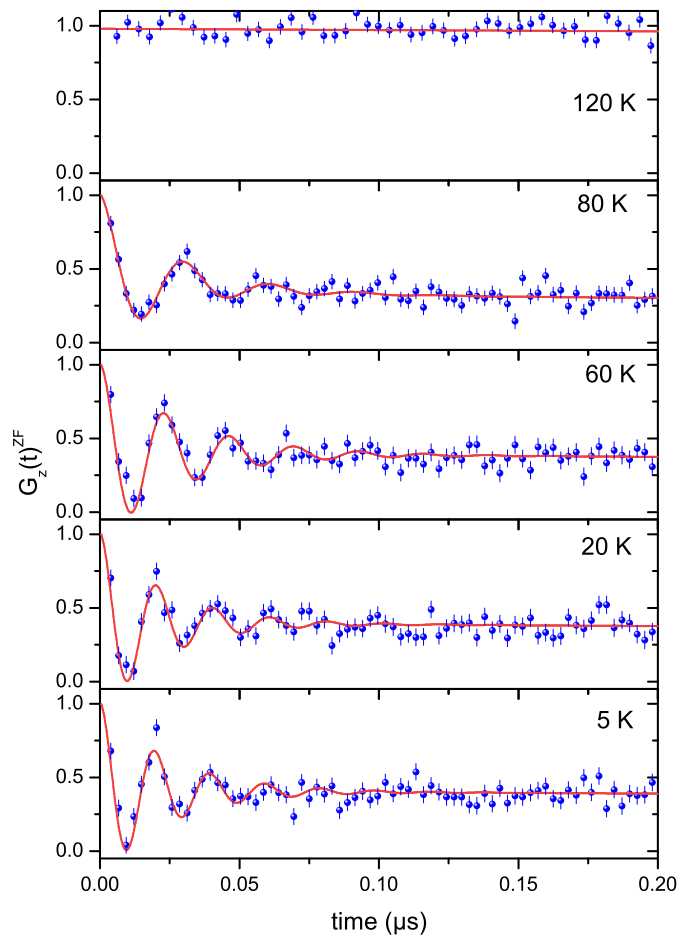}
\end{center}
\caption{ZF $\mu$SR patterns of La$_{2}$CoIrO$_{6}$ at some selected temperatures for a time window 0-0.2 $\mu$s, showing spontaneous muon spin rotation in the magnetically ordered state.}
\label{Fig_muSRZF0}
\end{figure}

The wTF results are supported by the ZF $\mu$SR measurements. The ZF spectra of $x$ = 0 at several temperatures are displayed in Fig. \ref{Fig_muSRZF0} for an early time window (for a wider time window see Fig. S2 in SM \cite{SM}). They show temperature dependent spontaneous rotation signals as expected for long-range magnetic order. These depolarization curves can be fitted with
\begin{equation}
G_{z}(t)^{ZF} = a_{lro} \cdot G_{z}(t)^{ZF}_{lro} + a_{ge} \cdot G_{z}(t)^{ZF}_{ge} + a^{ZF}_{BG}.
\label{Eq4}
\end{equation}
The first term stands for the signal from the long-range ordered volume with partial asymmetry weight $a_{lro}$:
\begin{equation}
G_{z}(t)^{ZF}_{lro} = \frac{2}{3}e^{-\lambda_{t}t}cos(2\pi\nu_{int}t)+\frac{1}{3}e^{-\lambda_{l}t},
\label{Eq5}
\end{equation}
where $\nu_{int}$ is the rotation frequency in local internal field caused by ordered electronic moments, $\lambda_{t}$ is the transverse damping parameter of the oscillations mainly caused by field inhomogeneities. The longitudinal damping $\lambda_{l}$ of the ``1/3 tail'' is related to field fluctuations along the direction of initial muon polarization. The $G_{z}(t)^{ZF}_{ge}$ = $e^{(-\lambda_{ge}t)^{\beta}}$ parameter is a so-called generalized exponential signal. For the present case of the $x$ = 0 compound, good fits can be achieved with an exponent $\beta$ = 1, representing a typical PM signal. $a_{lro}$ and $a_{ge}$ give the partial asymmetries for ordered and PM volume fractions. The undamped non-magnetic background contribution $a^{ZF}_{BG}$ has about 5\% of total asymmetry. 

The temperature dependence of $\nu_{int}$ follows a magnetization curve (Fig. S3 \cite{SM}). The transverse damping $\lambda_t$ of the rotating signal is hardly varying from 35(4) $\mu$s$^{-1}$ at 5 K to 40(5) $\mu$s$^{-1}$ at 80 K  and indicates a slight distribution of local fields. The longitudinal damping $\lambda_l$ can be kept zero below 80 K, typical for static order.

The ZF $\mu$SR also indicate the persistence of a PM volume fraction (about 10-15\% of total asymmetry) with finite damping varying between about 4  $\mu$s$^{-1}$ at 90 K to 0.8  $\mu$s$^{-1}$ at 5 K. One should note, however, that these values and those for the longitudinal damping of the ordered contribution are strongly correlated and cannot safely be separated. 

The FIM behavior of La$_{2}$CoIrO$_{6}$ was proposed to arise from the AFM coupling between Co$^{2+}$ and Ir$^{4+}$ by way of the hybridization of Co 3$d$ $t_{2g}$ and Ir 5$d$ $j_{eff}$ = 1/2 orbitals \cite{Noh}. From our XAS results, however, we conclude that there is present also a fraction of non-magnetic Ir$^{3+}$ in this compound not contributing to magnetic coupling. This, together with the frustration caused by the ASD at the Co/Ir sites, may serve as the origin for the PM volume fraction observed. While strong structural phase separation has to be excluded from SXRD, there could exist superparamagnetic regions with only small effective moment comprised within the ordered matrix. This could also explain that DC susceptibility does not show an additional PM signal, since the superparamagnetic contribution would be dominated by the ferrimagnetic response. Clearly more work is necessary to assess these observations.

For the Ca-doped samples, remarkable changes could be noticed compared with the parent compound. At this point it is necessary to comment on the way of analysis of the spectra for these compounds. As can be seen from the Figs. S4, S8 and S11 (SM \cite{SM}) of the ZF data a reliable ``true'' fit using Eq. \ref{Eq4} using several independent contributions is hardly feasible. This means these ``fits'' should be understood as heuristic ad hoc reproductions of the experimental data aiming to a reasonable systematic consistency. These qualitative data should be understood as preliminary since further experiments are necessary to allow a quantitative discussion of the damping behavior caused by fields from electronic and nuclear magnetic moments. Reliable information concerning magnetic and non-magnetic spectral contributions may, however, be traced from wTF experiments.

From wTF for $x$ = 0.5 we can see a continuous decrease of PM fraction starting around 100 K. At 90 K only about 15(3)\% of volume are long-range ordered. This is consistent with ZF data (Figs. S4, S5 \cite{SM}) which give an ordered fraction of 20(4)\% at this temperature (Fig. S6 \cite{SM}). At 60 K the PM fraction has decreased to about 30\%. The PM signal for $x$=0.5 has to be described by the generalized exponential depolarization with varying power parameter $\beta$, decreasing from about 0.80(5) at 90 K to 0.34(3) at 60 K. This is typical for a wide distribution of relaxation times as found with slowing down local field fluctuations in spin glasses \cite{Campbell}. At 30 K and below this fit strategy is not successful and instead of the generalized exponential we have to use a Dynamic Exponential Kubo-Toyabe (DEKT) function \cite{Uemura} close to static limit that is typical for a frozen spin glass with some residual fluctuations. 

Whilst the spontaneous frequencies of the ordered fraction cannot be well resolved at higher temperatures (due to low signal fraction), this is better possible at low temperatures and they are close to those found for the $x$ = 0 sample (see Fig. S3 \cite{SM}). This means that below 30 K we find a coexistence of a long-range ordered fraction (then about 40\% of total volume) similar to the undoped compound with a nearly statically frozen spin glassy volume.

For the $x$ = 0.8 sample we find from wTF a gradual partial loss of asymmetry (Fig. S7 \cite{SM}) below 100 K (corresponding to $T_{C1}$) followed by a plateau between 60 K (close to $T_{C2}$) and 40 K, and a drop below (corresponding to the freezing temperature of the SG state), \textit{i.e.} these transitions are in good agreement with the $M(T)$ and $\chi_{ac}$ data. ZF data are consistent with wTF. Spectra are shown in Fig. S8 \cite{SM}. The PM fraction is again best described by a generalized exponential damping function with $\beta$ about 0.5. Below the freezing temperature it turns to a nearly static DEKT signal of the SG. The long-range ordered fraction below the freezing temperature amounts to about 50(5)\% of sample. The spontaneous frequencies are reduced with respect to those found for $x$ = 0 and 0.5.

The loss of wTF asymmetry for the $x$ = 1.0 sample is continuous between 100 K and 50 K followed by a drop below (see Fig. S10 \cite{SM}). The ZF spectra are shown in Figs. S11 and S12 of SM \cite{SM}. The asymmetry of the ordered fraction stays roughly constant at about 15(3)\% of total asymmetry for all temperatures. Spontaneous frequencies may hardly be resolved. The power factor $\beta$ of the PM signal varies between 0.8 around 80 K to 0.5 at 30 K. At lower temperatures the general exponential signal again turns to a nearly static DEKT spin glass signal. 

The absence of a PM fraction at low temperatures for the Ca-doped samples as traced from the wTF data indicates a different coupling mechanism than that of the $x$ = 0 parent compound. The tendency toward non-magnetic Ir$^{5+}$ weakens the spin-selectivity of the hybridization between Co 3$d$ and Ir 5$d$ nearest neighbor orbitals, leading to the rise of AFM and the onset of SG. This in turn enhances the electronic transport, as will be discussed next.

\subsection{Electronic transport}

\begin{figure}
\begin{center}
\includegraphics[width=0.44 \textwidth]{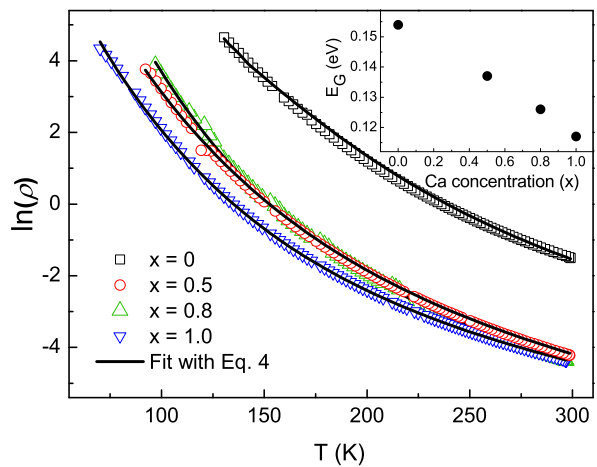}
\end{center}
\caption{DC electrical resistivity ($\rho$) as a function of temperature for La$_{2-x}$Ca$_{x}$CoIrO$_{6}$ samples. The solid lines represent the best fit of the data with Eq. \ref{Eq6}. The inset shows the evolution of $E_{G}$ with the Ca concentration, extracted from the fits.}
\label{Fig_rho}
\end{figure}

Fig. \ref{Fig_rho} displays the temperature dependent electrical resistivity ($\rho$) for the La$_{2-x}$Ca$_{x}$CoIrO$_{6}$ samples. The curves show insulating behavior for all compounds, and a clear decrease of $\rho$ for the doped samples relative to the parent compound. The $\rho$ of the doped samples are more than one order of magnitude smaller than that of $x$ = 0 parent compound. Although the Ca$^{2+}$ to La$^{3+}$ substitution is a hole doping, these results indicate that electrons are the charge carriers, since the increased Ca concentration signify a decrease in the number of localized 3d/5d electrons in the TM ions, which in turn reduce the Coulomb repulsion and favour the electrical conductance.

In resemblance to La$_{2-x}$Sr$_{x}$CoIrO$_{6}$ system \cite{Narayanan}, the ln($\rho$) didn`t show a linear dependence with $T^{-1}$ characteristic of a simple thermally activated mechanism of electronic transport nor the linear dependence with $T^{-1/4}$ of the variable-range hopping model \cite{Mott}. But the curves could be well fitted by a combination of both mechanisms:
\begin{equation}
\frac{1}{\rho}=\frac{1}{\rho_{01}}e^{-(E_{G}/k_{B}T)} + \frac{1}{\rho_{02}}e^{-(T_{0}/T)^{1/4}}, \label{Eq6}
\end{equation}
where $\rho_{01}$ and $\rho_{02}$ are pre-exponential factors, $E_{G}$ is the activated band gap and $k_{B}$ is the Boltzmann constant. The inset of Fig. \ref{Fig_rho} shows the decrease of $E_{G}$ with increasing Ca content, and Table \ref{T_rho} displays the main results obtained from the fittings.

The results described above clearly indicate that different mechanisms are contributing to the electronic transport in La$_{2-x}$Ca$_{x}$CoIrO$_{6}$. The data can be understood in terms of the structural and electronic changes induced by the introduction of Ca at La site. The lesser conductivity observed for the parent compound is manifested in the larger $\rho_{01}$ and $\rho_{02}$ values (Table \ref{T_rho}), which represent the expected resistivity at higher temperatures. The great decrease of $\rho$ observed from $x$ = 0 to $x$ = 0.5 might be related to the emergence of Co$^{3+}$, which would reduce the Coulomb repulsion and facilitates the hopping of the Ir $j$ = 1/2 electron to the Co $t_{2g}$ orbitals. 

It is interesting to note for $x$ = 0.8 sample that, although its $\rho$ values are similar to that of $x$ = 0.5 at high $T$, it becomes slightly larger as $T$ decreases. This is manifested in the $\rho_{02}$ values, reflecting the fact the second term of Eq. \ref{Eq6} dominates at lower temperatures. The XAS results indicated that from the $x$ = 0.5 to the 0.8 sample there is a tendency of increase in the proportion of Ir$^{5+}$ and a return to Co$^{2+}$ configuration. Thus, in one hand we have the increased Coulomb repulsion due to the introduction of electron at Co. On the other hand, the Ir$^{5+}$ state could facilitate the hopping of Co 3$d$ electrons via Ir empty orbitals. Furthermore, as discussed in previous sections the weaken hybridization of Co and Ir$^{5+}$ enhances the tendency of PM, and thus one see a subtle increase in $\rho$ of $x$ = 0.8 at lower $T$. For $x$ = 1.0, the Rietveld refinements showed a tendency of shrinkage of the lattice with increasing the Ca concentration, which would lead to stronger hybridization between the TM ions. This, together with the small ASD observed for $x$ = 0.8 sample, explain its slightly smaller $\rho$ values at high temperature. This is manifested in the smallest $T_0$ observed for this sample, since this parameter plays a role on the slope of the curves.

\begin{table}
\caption{Main results obtained from the fittings of $\rho$ $vs$ $T$ curves with Eq. \ref{Eq6}.}
\label{T_rho}
\begin{tabular}{c|cccc}
\hline \hline
Ca concentration (x) & $\rho_{01}$ ($\Omega$m) & $\rho_{02}$ ($\Omega$m)  & $T_{0}$ (K) & $E_{G}$ (eV) \\
\hline
0 & 6$\times$10$^{-4}$ & 1$\times$10$^{-7}$ & 2.3$\times$10$^{7}$ & 0.154 \\

0.5 & 9$\times$10$^{-5}$ & 1$\times$10$^{-9}$ & 2.7$\times$10$^{7}$ & 0.138 \\

0.8 & 1$\times$10$^{-4}$ & 2$\times$10$^{-5}$ & 2.3$\times$10$^{7}$ & 0.126 \\

1.0 & 2$\times$10$^{-4}$ & 4$\times$10$^{-9}$ & 2.9$\times$10$^{6}$ & 0.117 \\

\hline \hline
\end{tabular}
\end{table}

\begin{figure}
\begin{center}
\includegraphics[width=0.5 \textwidth]{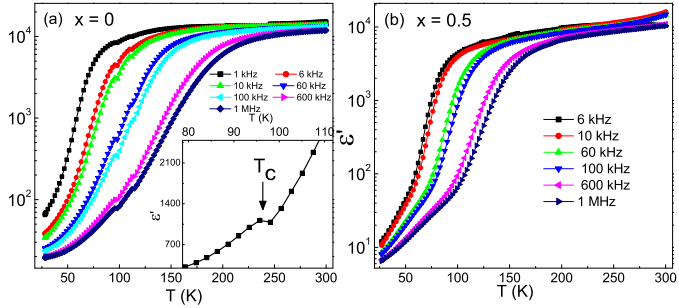}
\end{center}
\caption{Temperature dependence of the dielectric permittivity for (a) $x$ = 0 and (b) $x$ = 0.5 samples for some selected frequencies. The inset shows a magnified view of the $f$ = 30 kHz curve of $x$ = 0 sample, evidencing the anomaly around $T_C$.}
\label{Fig_Elinha}
\end{figure}

Due to their highly insulating behavior, the resistivity of the samples achieved the limit of detection of the equipment near the magnetic transition temperatures. To further investigate the electronic processes below the magnetic orderings, we measured AC dielectric permittivity from room temperature down to 27 K for two representative samples, $x$ = 0.0 and x = 0.5, at several frequencies ($f$) in the range 100 Hz - 1 MHz. Figs. \ref{Fig_Elinha}(a) and (b) show the $\varepsilon'$ curves for some selected $f$. For both samples it is observed a giant dielectric permittivity value ($\sim$10$^{4}$) near room temperature, followed by a $f$-dependent sharp drop of the curves, where the $\varepsilon'$ values at $f$ = 1 MHz decrease respectively to 20 and 7 for $x$ = 0 and 0.5 at 27 K. It is interesting to note an anomaly in the curves of the $x$ = 0 compound at $T$ corresponding to its magnetic ordering, as highlighted in the inset of Fig. \ref{Fig_Elinha}(a). This is a clear indicative of the correlation between the magnetic and the dielectric properties of this compound \cite{Song}. 

For $x$ = 0.5 such a kink is not observed, possibly because of its less intense and broader magnetic transition caused by changes in the Co/Ir valences and the enhanced ASD (see Fig. \ref{Fig_PPMS}). However, it can be noticed a clear change in the slope of the $\varepsilon'$ curves at $T$ close and below the magnetic transition temperature, with a linear dependence of log($\varepsilon'$) with $T$ at lower temperatures. This suggests that the coupling between magnetic and dielectric properties is also manifested in the Ca-doped samples.

\begin{figure}
\begin{center}
\includegraphics[width=0.5 \textwidth]{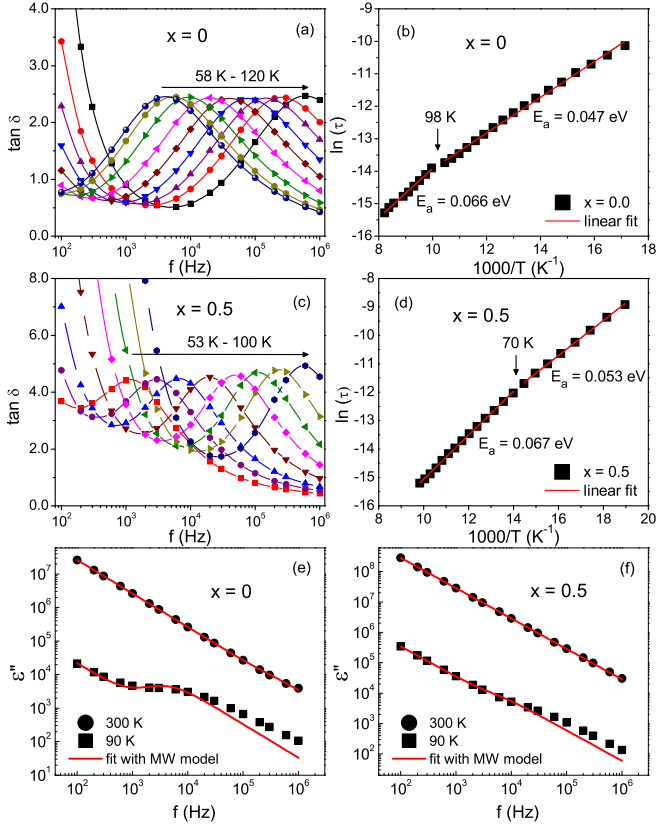}
\end{center}
\caption{(a) Variation in loss tangent (tan$\delta$) with frequency at different temperatures for the $x$ = 0 sample. (b) Relaxation time as a function of temperature for the $x$ = 0 sample. (c) Frequency dependence of tan$\delta$ for the $x$ = 0.5 sample at different temperatures. (d) Relaxation time as a function of temperature for the $x$ =0.5 sample. The $\varepsilon''$ $vs$ frequency curves for $x$ = 0 and 0.5 samples are displayed in (e) and (f), respectively. The red lines in (b) and (d) are the best fits with Arrhenius equation, while those in (e) and (f) are the best fits with the MW model.}
\label{Fig_E2linhas}
\end{figure}

Fig. \ref{Fig_E2linhas}(a) shows the frequency variation of dielectric loss tangent (tan$\delta$) for $x$ = 0 sample at temperatures in the 58-128 K range. Only a single Debye-like peak with a frequency maximum ($f_{max}$) was observed in this figure. The tan$\delta$ peak indicates dielectric relaxations in this sample within the measured frequency and temperature ranges. The tan$\delta$ peak shifting toward higher $f$ with increasing temperature indicate that the dielectric relaxation is thermally activated. Thus, the relaxation times ($\tau$) follow an Arrhenius equation given as $\tau$ = $\tau_{0}e^{(E_a/k_B T)}$, where $\tau_0$ is the relaxation time at $T$ $\to$ $\infty$, $E_a$ is the activation energy, $k_B$ is the Boltzmann constant and $T$ is the absolute temperature. Activation energies are calculated from the plot drawn between $ln(\tau$) against 1000/$T$, as shown in Fig. \ref{Fig_E2linhas}(b), where the relaxation time $\tau$=(2$\pi f_{max})^{-1}$ was determined for each temperature. From linear slopes in Fig. \ref{Fig_E2linhas}(b) the obtained activation energies are 0.047 and 0.066 eV at temperatures below and above 98 K, respectively.

The tan$\delta$ as a function of $f$ for $x$ = 0.5 sample is shown in Fig. \ref{Fig_E2linhas}(c) in the 53-100 K temperature range. A similar dielectric relaxation is also observed for this sample. Its temperature dependence of relaxation time is shown in Fig. \ref{Fig_E2linhas}(d). Two distinct relaxation regions were also observed for this sample below and above 70 K. The calculated activation energies for these regions were 0.053 and 0.067 eV. These activation energies are close to those obtained for the $x$ = 0 sample, indicting that the relaxation mechanism is basically the same in both samples.

Impedance measurements evidence a well-defined dielectric relaxation for $x$ = 0 sample, while for $x$ = 0.5 two distinct contributions were found, attributed to relaxations of the grains and of the grain boundaries (see SM \cite{SM}). For both samples is was observed distinct relaxation processes at low and high temperatures, endorsing the dielectric permittivity and the $ac$ conductivity results. The activation energies obtained from the $ac$ conductivity measurements at low and high temperatures were similar to those acquired from the dielectric permittivity, indicating that the carriers responsible for the dielectric relaxation and electrical conductivity are fundamentally the same \cite{SM}.

The unconventional dielectric features described above resemble those of magnetodielectric La$_{2}$CoMnO$_{6}$ and La$_{2}$NiMnO$_{6}$ DPs \cite{Murthy3,Sarma,Kumar}. A common ground of these systems is the centrosymmetric $P2_{1}/n$ structure, which in principle precludes the possibility of relative displacement of cations and anions. The dielectric responses of the above mentioned perovskites were interpreted in terms of the breaking of inversion symmetry caused by charge ordering of Co$^{2+}$/Ni$^{2+}$ and Mn$^{4+}$. In the case of the compounds here investigated, it could arise from Co/Ir charge ordering and/or it could be due to dipolar contributions originating from the asymmetric hopping of charge carriers between Co$^{2+}$ and Ir$^{4+}$ in the presence of an electric field, which is expected to give rise to a Debye relaxation \cite{Kao}. However, the behavior here observed for $x$ = 0 and 0.5 samples in the $f$ range of the experiments can also originate from extrinsic sources as Maxwell-Wagner (MW) relaxations, \textit{i.e.}, it can be due to the presence of accumulated charge carriers between regions in the sample that have different conductivities such as near the grain boundaries \cite{Hippel}.  

The MW polarization mechanism can be identified by its characteristic $f^{-1}$ dependence of the imaginary dielectric permittivity at lower $f$, whereas the Debye part of $\varepsilon''$ data goes to zero \cite{Hippel}. Figs. \ref{Fig_E2linhas}(e) and (f) show the log($\varepsilon''$) $vs$ log($f$) curves at some selected temperatures for both $x$ = 0 and 0.5 samples. As can be noticed, at lower frequencies the curves are linear with slope close to -1, clearly suggesting the presence of MW relaxations in these partially disordered compounds. In fact, at high $T$ the curves could be well fitted with the MW model. However, for $T$ close and below the magnetic ordering the MW model fails to describe the dielectric relaxation at higher $f$. As can be noticed in Figs. \ref{Fig_E2linhas}(e) and (f) the 90 K curves can be well fitted by the MW model only for $f<$ 10 kHz, suggesting that an additional relaxation process is present, possibly a Debye-like relaxation process arising from an asymmetric hopping between the TM sites in the presence of the applied electric field \cite{Song,Sarma}. This mechanism would be highly spin dependent, since this hopping depends on the relative-spin orientations between neighboring magnetic ions, being thus both investigated samples prospective candidates to exhibit magnetodielectric effect.
 
\section{Summary}

In this work we thoroughly investigated the structural, electronic and magnetic properties of La$_{2-x}$Ca$_{x}$CoIrO$_{6}$ ($x$ = 0, 0.5, 0.8, 1.0) samples by means of several experimental techniques. Our XAS and XMCD results at Co-$K$ and Ir-$L_{2,3}$ edges indicate the increased amount of Co$^{3+}$ for $x$ = 0.5 when compared to $x$ = 0 parent compound, while for larger Ca-doping is observed a tendency toward Ir$^{5+}$ state, which is corroborated by the SXRD and Raman spectroscopy data. Such changes affect the magnetic coupling between the TM ions, where the FIM behavior resulting from the presence of  FM sublattices at Co and Ir sites observed for $x$ = 0 seems to weaken and evolve to AFM for the doped samples. The competing magnetic interactions resulting from the mixed valence states lead to SG state coexisting with an ordered fraction in the doped compounds, as shown by the AC and DC magnetization results. Our $\mu$SR results show remarkable changes from the $x$ = 0 parent compound to the Ca-doped materials. For the parent compound there is observed the persistence of a PM volume fraction down to the lowest investigated temperature, 5 K, ascribed to the presence of Ir$^{3+}$ and ASD. For the Ca-doped ones, there is observed the freezing of the PM phase below 30 K, in agreement with the AC magnetization data.  Investigation of electronic transport shows a highly insulating behavior for all samples and a possible magnetodielectric effect in $x$ = 0 and 0.5 compounds.

\begin{acknowledgements}
This work was supported by Conselho Nacional de Desenvolvimento Cient\'{i}fico e Tecnol\'{o}gico (CNPq) [No. 400134/2016-0], Funda\c{c}\~{a}o Carlos Chagas Filho de Amparo \`{a} Pesquisa do Estado do Rio de Janeiro (FAPERJ), Funda\c{c}\~{a}o de Amparo \`{a} Pesquisa do Estado de Goi\'{a}s (FAPEG), Funda\c{c}\~{a}o de Amparo \`{a} Pesquisa do Estado de S\~{a}o Paulo (FAPESP) and Coordena\c{c}\~{a}o de Aperfei\c{c}oamento de Pessoal de N\'{i}vel Superior (CAPES).  The authors thank the DXAS and XPD staffs of LNLS for technical support and LNLS for the concession of beam time (proposals No. 20160578 and No. 20170057). Experimental support at Paul Scherrer Institut by H. Luetkens is gratefully acknowledged. E.S., F.J.L. and E.B.S. acknowledge support by a joint DFG-FAPERJ project Li 244/12. F.J.L. is grateful for a fellowship by FAPERJ.
\end{acknowledgements}

\end{document}